\RequirePackage{lineno}
\documentclass[%
 reprint,
superscriptaddress,
 amsmath,amssymb,
 aps,
]{revtex4-2}

\usepackage{graphicx}   
\usepackage{placeins}
\usepackage{xcolor}
\usepackage{natbib}
\usepackage{hyperref}
\usepackage{amsmath,amssymb,amsfonts}
\usepackage{xspace}
\usepackage{subfigure}
\usepackage{soul}

\newcommand{\sevenn}        {$\sqrt{s_{\mathrm{NN}}}~=~7$~TeV\xspace}

\newcommand{\xii}{$\Xi$\xspace}
\newcommand{\alam}{$\overline{\Lambda}$\xspace}
\newcommand{\lam}{$\Lambda$\xspace}
\newcommand{\ks}{$\mathrm{K}^{0}_{\mathrm S}$\xspace}

\newcommand{\om}{$\Omega$\xspace}

\newcommand{\ppt}{\ensuremath{p_{\rm T}}\xspace}

\newcommand {\auau}  {\ensuremath{\rm Au+Au}\xspace}
\newcommand {\pbpb}  {\ensuremath{\rm Pb+Pb}\xspace}
\newcommand {\xexe}  {\ensuremath{\rm Xe+Xe}\xspace}
\newcommand {\oo}  {\ensuremath{\rm O+O}\xspace}
\newcommand {\pp}    {\ensuremath{\rm p+p}\xspace}
\newcommand {\ppb}    {\ensuremath{\rm p+Pb}\xspace}
\newcommand {\xxi}       {$\Xi^{-}+\overline{\Xi}^{+}$\xspace}
\newcommand {\oom}       {$\Omega^- + \overline{\Omega}^+$\xspace}
\newcommand {\dndeta}       {\ensuremath{\mathrm{d}N_\mathrm{ch}/\mathrm{d}\eta}\xspace}

\newcommand {\avdndeta}     {\ensuremath{\langle\dndeta\rangle}\xspace}

\setstcolor{red}

\usepackage[bottom]{footmisc}

\begin{document}

\title{Multiplicity dependence of (multi)strange hadrons in oxygen-oxygen collisions at \sevenn using EPOS4 and AMPT}

\author{M.~U.~Ashraf}
\email{muashraf@wayne.edu}
\affiliation{Department of Physics and Astronomy, Wayne State University, 666 W. Hancock, Detroit, Michigan 48201, USA}%

\author{A.~M.~Khan}
\email{akhan129@gsu.edu}
\affiliation{Georgia State University, Atlanta, GA 30303, USA}%

\author{J.~Singh}
\email{jsingh2@bnl.gov}
\affiliation{Instituto de Alta Investigaci\'on, Universidad de Tarapac\'a, Casilla 7D, Arica, 1000000, Chile\\
}

\author{G.~Nigmatkulov}
\email{gnigmat@uic.edu}
\affiliation{Department of Physics, University of Illinois Chicago, Chicago, IL 60607, USA}

\author{H.~Roch}
\email{hendrik.roch@wayne.edu}
\affiliation{Department of Physics and Astronomy, Wayne State University, 666 W. Hancock, Detroit, Michigan 48201, USA}

\author{S.~Kabana}
\email{sonja.kabana@cern.ch}
\affiliation{Instituto de Alta Investigaci\'on, Universidad de Tarapac\'a, Casilla 7D, Arica, 1000000, Chile\\
}

\date{\today}

\begin{abstract}

It is anticipated that the Large Hadron Collider (LHC) will collect data from oxygen-oxygen (\oo) collisions at a center-of-mass energy of \sevenn to explore the effects observed in high multiplicity proton-proton (\pp) and proton-lead (\ppb) collisions that are closely related to lead-lead (\pbpb) collisions.
These effects include azimuthal asymmetries in particle production and variations in the abundances and momentum distributions across different hadron species, which are indicative of collective particle production mechanisms induced by the interactions in the presence of a QGP.  The upcoming data on \oo collisions at the LHC are expected to constrain the model parameters and refine our understanding of theoretical models. In this work, the predicted transverse momentum (\ppt) spectra, rapidity density distributions ($\mathrm{d}N/\mathrm{d}y$), particle yield ratios, and \ppt-differential ratios of (multi)strange hadrons produced in \oo collisions at \sevenn using AMPT and EPOS4 models are presented. 
AMPT focuses on partonic and hadronic interactions, while EPOS4 incorporates hydrodynamics.
Stronger radial flow in EPOS4 is also observed compared to AMPT. 
Both models predict the final-state multiplicity overlap with \pp, \ppb, and \pbpb collisions. 
\end{abstract}

\maketitle

\section{Introduction}
\label{sec1}
High-energy heavy-ion collisions provide an opportunity to create and study the Quark-Gluon Plasma (QGP) --- a strongly interacting state of matter that is described by Quantum Chromodynamics (QCD)~\cite{1, 2, 3}. In recent years, there has been a lot of debate about the limit of the smallest possible droplet of the QGP that could be formed in collisions. Experiments at the Relativistic Heavy Ion Collider (RHIC) and the Large Hadron Collider (LHC) have facilitated investigations of the properties of the QGP. Collisions of symmetric heavy ions, such as lead-lead (\pbpb) and gold-gold (\auau), in these facilities have revealed that the QGP exhibits hydrodynamic behavior, flowing like a nearly perfect fluid with a small shear-viscosity-to-entropy-density ratio~\cite{21, 22}. This interpretation of nucleus-nucleus ($\mathrm{A+A}$) results depends on the comparison with the results from small collision systems, such as proton-proton (\pp) or proton-nucleus ($\mathrm{p+A}$) because it allows us to identify and understand the nuclear effects and final-state phenomena in large systems.

\begin{table*}[!t]
    \centering
    \caption{Average charged-particle multiplicity $\langle \dndeta \rangle$ at $|\eta| < 0.5$ and $\langle N_{\rm part} \rangle$ in \oo collisions at \sevenn for different centrality classes using AMPT-Def, AMPT-SM and EPOS4.}
    \begin{tabular}{|c|c|c|c|c|c|c|}
    \toprule
    Centrality ($\%$) & \multicolumn{2}{c|}{AMPT-Def}   & \multicolumn{2}{c|}{AMPT-SM} & \multicolumn{2}{c|}{EPOS} \\ & 
     $\ensuremath{\langle\mathrm{d}N_\mathrm{ch}/\mathrm{d}\eta}\rangle\xspace\pm{\rm rms}$ & $\langle{N_{\rm part}}\rangle\pm{\rm rms}$ & $\ensuremath{\langle\mathrm{d}N_\mathrm{ch}/\mathrm{d}\eta}\rangle\xspace\pm{\rm rms}$ & $\langle{N_{\rm part}}\rangle\pm{\rm rms}$ & $\ensuremath{\langle\mathrm{d}N_\mathrm{ch}/\mathrm{d}\eta}\rangle\xspace\pm{\rm rms}$ & $\langle{N_{\rm part}}\rangle\pm{\rm rms}$\\
    \hline   
            $0$~--~$5$    & $185.556\pm0.016$ &$28.82\pm2.12$ & $188.293\pm0.043$ &$29.00\pm2.03$ & $236.440\pm0.142$ &$27.86\pm2.18$\\
            $5$~--~$10$   & $142.550\pm0.015$ &$26.35\pm2.69$ & $145.678\pm0.038$ &$26.78\pm2.60$ & $189.801\pm0.130$ &$25.05\pm2.33$\\
            $10$~--~$20$  & $108.771\pm0.009$ &$22.77\pm3.28$ & $110.442\pm0.023$ &$23.24\pm3.24$ & $148.437\pm0.083$ &$21.44\pm3.15$\\
            $20$~--~$30$  & $75.674\pm0.008$ &$18.13\pm3.29$ & $76.085\pm0.019$ &$18.54\pm3.26$ & $105.863\pm0.071$ &$16.87\pm3.12$\\ 
            $30$~--~$40$  & $51.864\pm0.006$ &$14.03\pm3.08$ & $51.471\pm0.016$ &$14.33\pm3.05$ & $75.027\pm0.064$ &$12.44\pm2.56$\\ 
            $40$~--~$50$  & $34.789\pm0.005$ &$10.58\pm2.79$ & $34.188\pm0.013$ &$10.80\pm2.75$ & $53.037\pm0.050$ &$9.60\pm2.52$\\ 
            $50$~--~$60$  & $22.701\pm0.004$ &$7.84\pm2.42$ & $22.082\pm0.010$ &$7.99\pm2.40$  & $37.146\pm0.062$ &$6.98\pm2.30$\\ 
            $60$~--~$80$  & $11.493\pm0.002$ &$4.86\pm1.99$ & $11.117\pm0.005$ &$4.94\pm2.00$  & $19.889\pm0.023$ &$4.49\pm1.96$\\ 
            $80$~--~$100$ & $3.860\pm0.001$ &$2.60\pm1.01$  & $3.761\pm0.003$ &$2.63\pm1.03$  & $5.127\pm0.011$ &$1.89\pm1.53$\\ 
    \hline
    \hline
    \end{tabular}
    \label{tab1}
\end{table*}

Strangeness enhancement has been widely recognized and observed as a key signature of the QGP, as the hot medium facilitates the thermal production of strange quarks~\cite{4}.

The ratios of production yields for various strange hadrons relative to pions in $\mathrm{A+A}$ collisions from SPS to LHC show significant centrality and energy dependence~\cite{17}.
Recent studies~\cite{5, 6} at the LHC have observed features such as azimuthal anisotropies, modified hadron yields, and spectra in high-multiplicity $\mathrm{p+p}$ and \ppb collisions that are surprisingly similar to those observed in $\mathrm{A+A}$ collisions, despite the significant difference in system size~\cite{ALICE:2019zfl, 5}. These findings also exhibit qualitative agreement with predictions from the statistical hadronization model (SHM)~\cite{60}. The model describes particle production in terms of thermal equilibrium without QGP formation~\cite{Vovchenko:2019kes, ALICE:2018pal, 61, 62}. 

These findings are also consistent with the core-corona model~\cite{37, 63, 64, Kanakubo:2019ogh, Kanakubo:2021qcw} based on the assumption that strange quarks are produced thermally in the core, a high-density region of the colliding nuclei. On the other hand, the most commonly used Monte Carlo (MC) models for \pp collisions like Pythia~\cite{65, 66} and EPOS-LHC~\cite{56} are unable to quantitatively describe the strangeness enhancement observed in existing experimental data~\cite{5}. 
These observations present a significant challenge to the current theoretical understanding of QGP formation, as they suggest the possibility of QGP-like behavior in small collision systems. Consequently, to understand the enhanced production of strange hadrons in small systems, experimental investigation and advancement in hydrodynamic and transport models are required. The hydrodynamic and transport approaches complement each other. Hydrodynamic models provide direct access to the equation of state (EoS) and transport coefficients, while transport models can address non-equilibrium dynamics and provide a microscopic picture of the interactions~\cite{Lin:2021mdn}.

To further understand QGP formation mechanisms in small systems, LHC experiments are anticipated to collect data from oxygen-oxygen (\oo) collisions at \sevenn~\cite{23, 24}. This offers a significant and timely opportunity to investigate these effects in a system with a small number of participating nucleons and a final-state multiplicity comparable to that of smaller systems but with a larger geometrical transverse overlap. Strangeness production studies in \oo collisions, in a multiplicity range that bridges \pp and \ppb on the low side, and \pbpb and \xexe on the high side~\cite{24}, are of primary interest for determining the hadrochemistry of the formed medium and for studying hadronization. The underlying mechanisms of particle production in \oo collisions have recently been explored in several theoretical studies~\cite{25, 26, 27, 28, 29, 30, 31, 32, 33, 34, Bashir:2025hyt, Singh:2025pti}.  

The predictions presented in this work are based on the recently updated versions of the EPOS (EPOS4) and AMPT models. EPOS4 reasonably reproduces the overall strangeness enhancement observed in the ALICE data~\cite{42}. The version (v1.26t9b-v2.26t9b) of the AMPT model used in this study does not provide a consistent description of the yields and $p_{\rm T}$ spectra of multistrange baryons in heavy-ion collisions.~\cite{Shao:2020sqr}. Both models employ different mechanisms for strangeness production. EPOS4 includes the QGP phase in its hydrodynamic evolution with the lattice QCD EoS. AMPT, on the other hand, does not have a fully chemically equilibrated system compared to EPOS4. The partonic stage in AMPT is described by a parton cascade where interactions are primarily 2-to-2 elastic scatterings. In AMPT, the system can be isotropic in momentum space but not chemically equilibrated. 
In this article, we report on the predictions of various strange hadrons (\ks, {\lam (\alam)}, {\xxi}, {\oom}, and $\phi$) produced in \oo collisions at \sevenn using the AMPT~\cite{67, 68} and EPOS4~\cite{39, 40, 42} models. The observables under study include \ppt spectra, rapidity density distributions ($\mathrm{d}N/\mathrm{d}y$), multiplicity dependence of the yield ratios of strange hadrons relative to pions, and \ppt-differential strange-baryon-to-meson ratios. The results reported here are based on the preliminary study reported in Ref.~\cite{Singh:2025qab}, offering a more detailed investigation and broader set of observables.

The paper is organized as follows. Section~\ref{sec2} briefly presents event generation methodology with the AMPT and EPOS4 models. The results are reported and discussed in Section~\ref{sec3}. The summary, primary findings, and potential outlook for future research are given in Section~\ref{sec4}.  

\section{Event Generators}
\label{sec2}

The section provides a short description of the AMPT and EPOS4 models. The multiplicity classes in these predictions are based on the pseudorapidity distribution of charged particles within the region $|\eta| < 0.5$, following the same approach used in the experimental data~\cite{ALICE:2018pal}. The corresponding $\langle \dndeta \rangle$ and $\langle N_{\rm part} \rangle$ values for different centrality classes for both models are given in Table~\ref{tab1}. 

\subsection{AMPT}
The AMPT model was developed to study the dynamics of relativistic heavy-ion collisions and has been extensively used for various studies at RHIC and LHC energies~\cite{67, 68}. Both default (AMPT-Def) and string melting (AMPT-SM) versions have been used in this work. In AMPT, the HIJING model~\cite{Wang:1991hta} provides the initial spatial and momentum distributions of minijet partons and soft string excitations. The subsequent space-time evolution of these partons is then governed by the ZPC parton cascade model~\cite{Zhang:1997ej}. Following the parton cascade evolution, the model transitions the remaining partonic degrees of freedom into final-state hadrons using either string fragmentation or quark coalescence mechanisms. Subsequently, the interactions of these newly formed hadrons are governed by A Relativistic Transport (ART) model~\cite{Li:1995pra}. 

The default version of the AMPT model reasonably reproduces the rapidity distributions and \ppt spectra of identified particles at SPS and RHIC energies. This is due to the fact that it only involves minijet partons from HIJING in the parton cascade and uses the Lund string fragmentation for the hadronization~\cite{65}. However, the default version significantly underestimates elliptic flow at RHIC energies~\cite{Lin:2001zk}. 
In AMPT-SM, partonic and hadronic interactions are included, as well as the transition between these two phases~\cite{Lin:2001zk}. A simple quark coalescence model is then employed to convert these quarks back into hadrons after the ZPC. This approach has been successful in describing anisotropic flow in both large and small collision systems~\cite{Lin:2001zk, Bzdak:2014dia, Ma:2014pva}. 

In this study, we analyzed $\sim 6$ million minimum-bias events for both AMPT-Def and AMPT-SM.

\subsection{EPOS4}
EPOS4 is a multipurpose event generator that uses a unique approach to treat all collision systems (\pp, $\rm p+A$ and $\mathrm{A+A}$).
The EPOS model simulates nucleus-nucleus ($\mathrm{A+A}$) collisions using a 3+1 dimensional viscous hydrodynamic approach~\cite{37}. The initial conditions in EPOS4 are defined in terms of flux tubes using the Gribov-Regge multiple scattering theory~\cite{38}. The core-corona, hydrodynamical evolution, and the hadronic cascade are the main components of EPOS4. The fragmentation of flux tubes into core and corona (which later hadronize into hadron jets) is determined by the probability of a fragment escaping the bulk matter. This probability depends on the transverse momentum of a fragment and the local string density. EPOS4~\cite{42} further incorporates a mechanism to distinguish between the core and corona zones of particle production within the colliding nuclei. To account for the density variations within the system, Regge theory~\cite{38} is used for the low density (corona), and hydrodynamic equations are used for the high density (core). The core utilizes a Cooper-Frye procedure to transition from the fluid to the particles. The vHLLE algorithm~\cite{Allton:2002zi}, a viscous hydrodynamic approach implemented in a 3+1D framework, is employed with an EoS derived from lattice QCD. The hadronic afterburner, a hadronic cascade model based on the UrQMD model~\cite{Bleicher:1999xi, Bass:1998ca}, is employed to describe the late stage hadronic rescatterings and decays~\cite{Bleicher:1999xi, Bass:1998ca, Petersen:2008dd}. 

The core-corona model reproduces many features of \pp and \pbpb collisions~\cite{42}. Including the core-corona distinction improved the description of the centrality dependence of resonance production and strange particle production in $\mathrm{A+A}$ collisions~\cite{Knospe:2015nva}.
Additional details regarding the EPOS4 version used in this work can be found in Ref.~\cite{42}.

A total of $\sim 3$ million minimum-bias events were simulated from EPOS4. The simulations were conducted with specific parameters. To incorporate core-corona effects, the ``full'' core option was activated. Furthermore, the ``hydro'' parameter was enabled to model the hydrodynamic evolution, describing the collective particle flow. The EoS was the standard 3-flavor crossover EoS that is consistent with the lattice QCD calculations at zero net density~\cite{Werner:2013tya}. The hadronic cascade phase for interactions between particles after the initial collision was enabled.

\section{Results and Discussion}
\label{sec3}

\begin{figure*}[t!]
    \centering
    \begin{subfigure}
        \centering
        \includegraphics[width=0.45\linewidth]{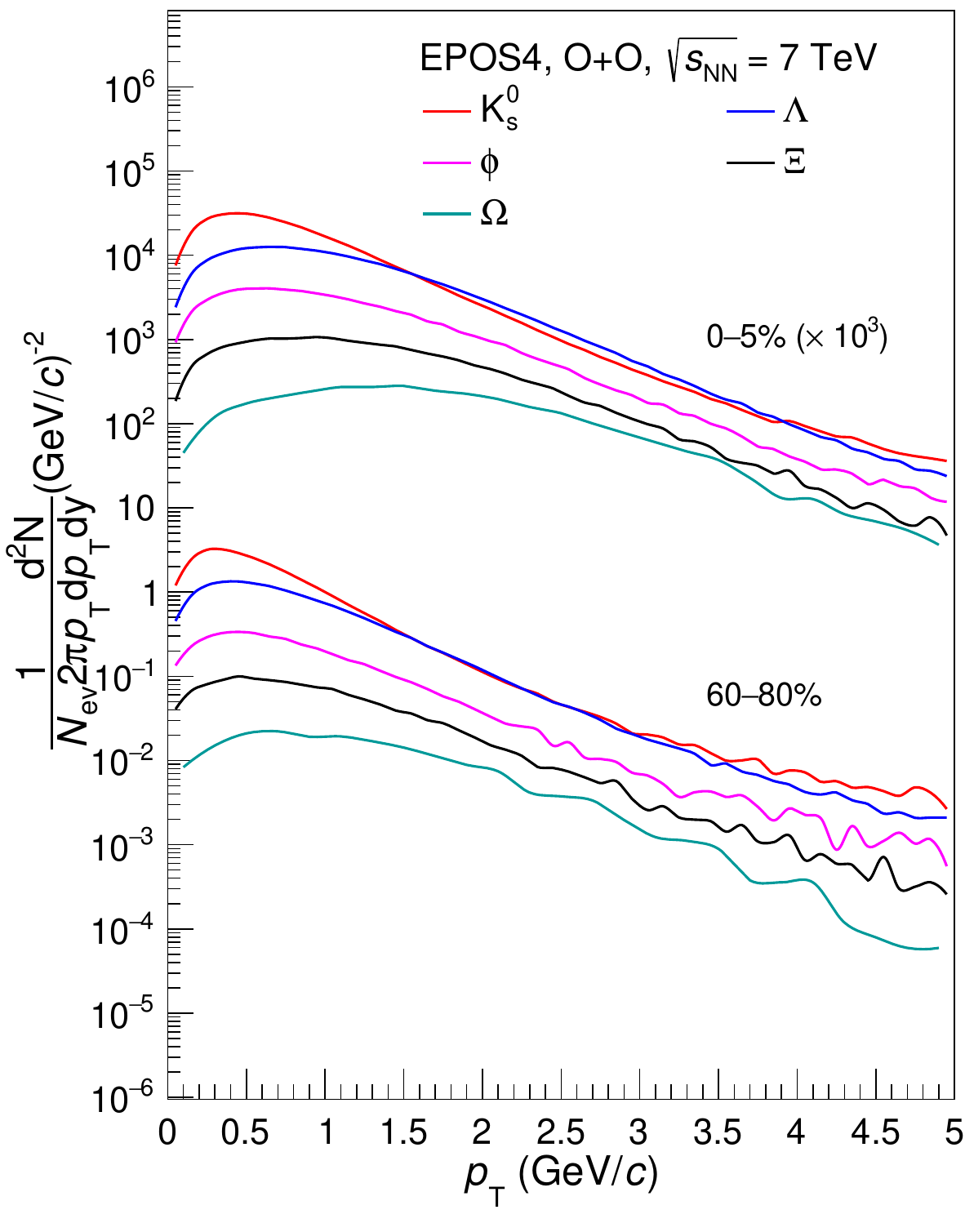} 
        \includegraphics[width=0.45\linewidth]{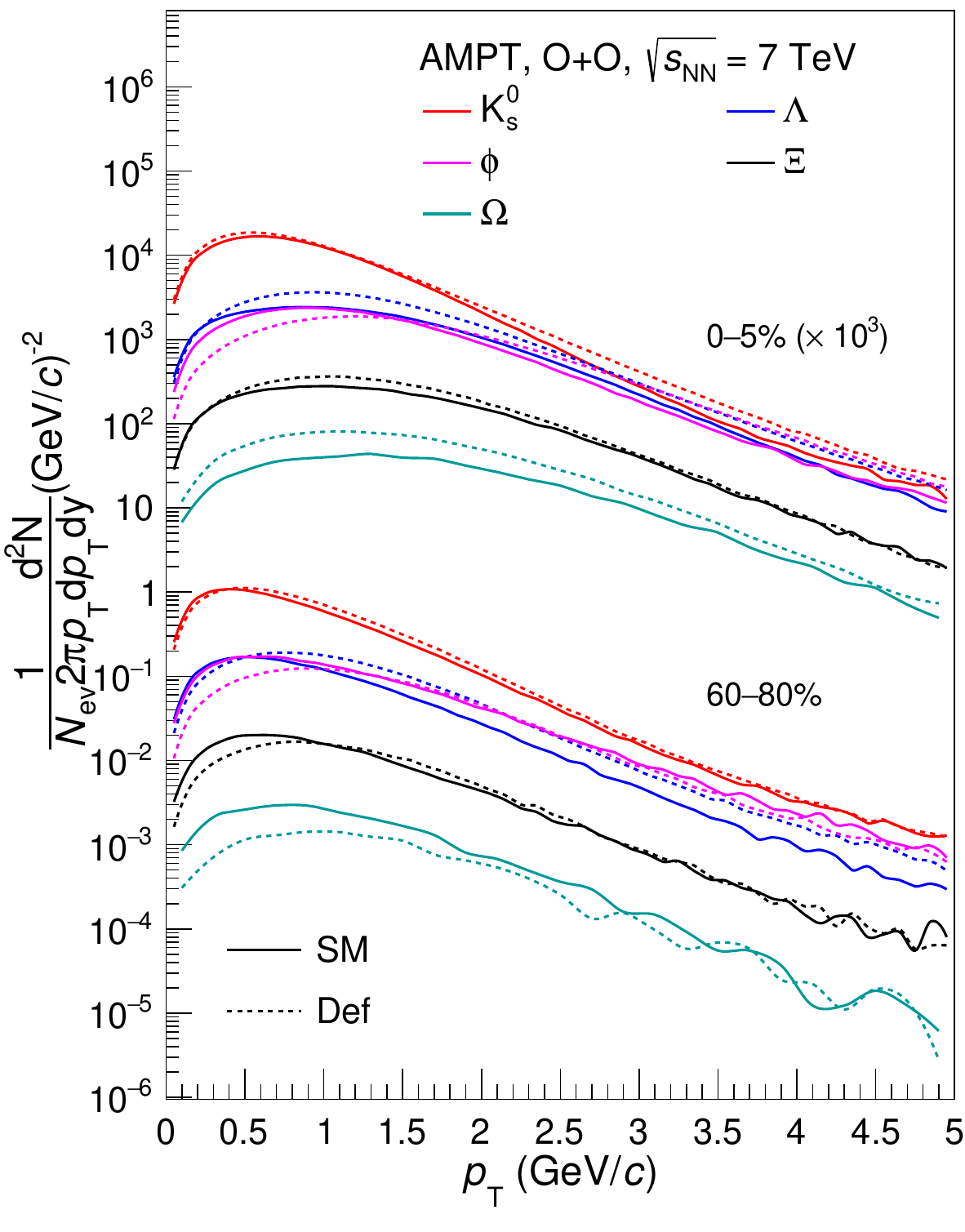}
    \end{subfigure}%
    \caption{(Color online) \ppt-spectra of (multi)strange hadrons ({\ks}, $\Lambda$, $\Xi$, $\phi$ and $\Omega$) from EPOS4 (left) and  AMPT (right) in \oo collisions at \sevenn for most central (0--5\%) and most peripheral (60--80\%) centrality classes. 0--5\% \ppt spectra for (multi)strange hadrons are scaled by a factor of $10$ for better visualization. In the right-side figure, solid lines represent AMPT-SM, while dotted lines represent AMPT-Def.}
    \label{fig1}
\end{figure*}

The transverse momentum (\ppt) distributions of {\ks}, $\Lambda$, $\Xi$, $\phi$, and $\Omega$ from AMPT-Def, AMPT-SM, and EPOS4 in the most central (0--5\%) and most peripheral (60--80\%) \oo collisions at \sevenn are shown in Fig.~\ref{fig1}. The EPOS4 results are depicted by solid lines in Fig.~\ref{fig1} (left). In contrast, the AMPT-Def and AMPT-SM model results are represented by dotted and solid lines, respectively, in Fig.~\ref{fig1} (right).

The \ppt distributions of different particle species exhibit a clear evolution with centrality as well as particle type. At low \ppt, the distributions tend to flatten, with this effect being stronger for heavier particles than lighter ones, demonstrating a mass ordering behavior. This is expected in hydrodynamic models as a consequence of the blue shift induced by the collective expansion of the system. At intermediate \ppt, the spectra of heavier particles seem to converge with those of lighter particles. This may be attributed to radial flow. As the system expands, the radial flow can impart additional momentum to particles with lower \ppt, pushing them towards the intermediate-\ppt region~\cite{43, 44}. The slope of the \ppt-spectra generally becomes flatter with the increase in centrality, which is evident from Fig.~\ref{fig1}. In 0--5\% central collisions, the decrease in the slope of \ppt spectra is more pronounced compared to 60--80\% peripheral collisions, indicating the presence of stronger radial flow. Furthermore, heavier particles exhibit a significantly flatter slope compared to lighter particles, consistent with increasing radial flow for increasing particle mass. A similar trend has been observed in the \ppt spectra of identified hadrons in \oo collisions using EPOS4~\cite{34}.

The production of strange hadrons at low \ppt in AMPT and EPOS4 is driven by different mechanisms. The shape of the spectra is more influenced by partonic scatterings and hadronic rescatterings in AMPT, while hydrodynamic evolution plays a dominant role in EPOS4. As a result, EPOS4 exhibits stronger radial flow than AMPT. At intermediate \ppt, EPOS4 incorporates collective flow, viscous hydrodynamics, and fragmentation-recombination hadronization, generally producing smoother \ppt spectra, while AMPT-SM incorporates only quark coalescence. In the high-\ppt region, particles in AMPT are primarily produced through minijets or jets, followed by hadronization. The treatment of jet energy loss in AMPT can influence the suppression of high-\ppt hadrons. In contrast, EPOS4 considers core-corona separation, which affects the high-\ppt region of the spectra. The corona region (non-hydrodynamic) generates high-\ppt particles via jet fragmentation, while the core region produces lower \ppt particles through hydrodynamics. Consequently, EPOS4 may exhibit slightly different suppression patterns for high-\ppt hadrons as compared to AMPT.
 
\begin{figure}[!ht]
\includegraphics[width=0.45\textwidth]{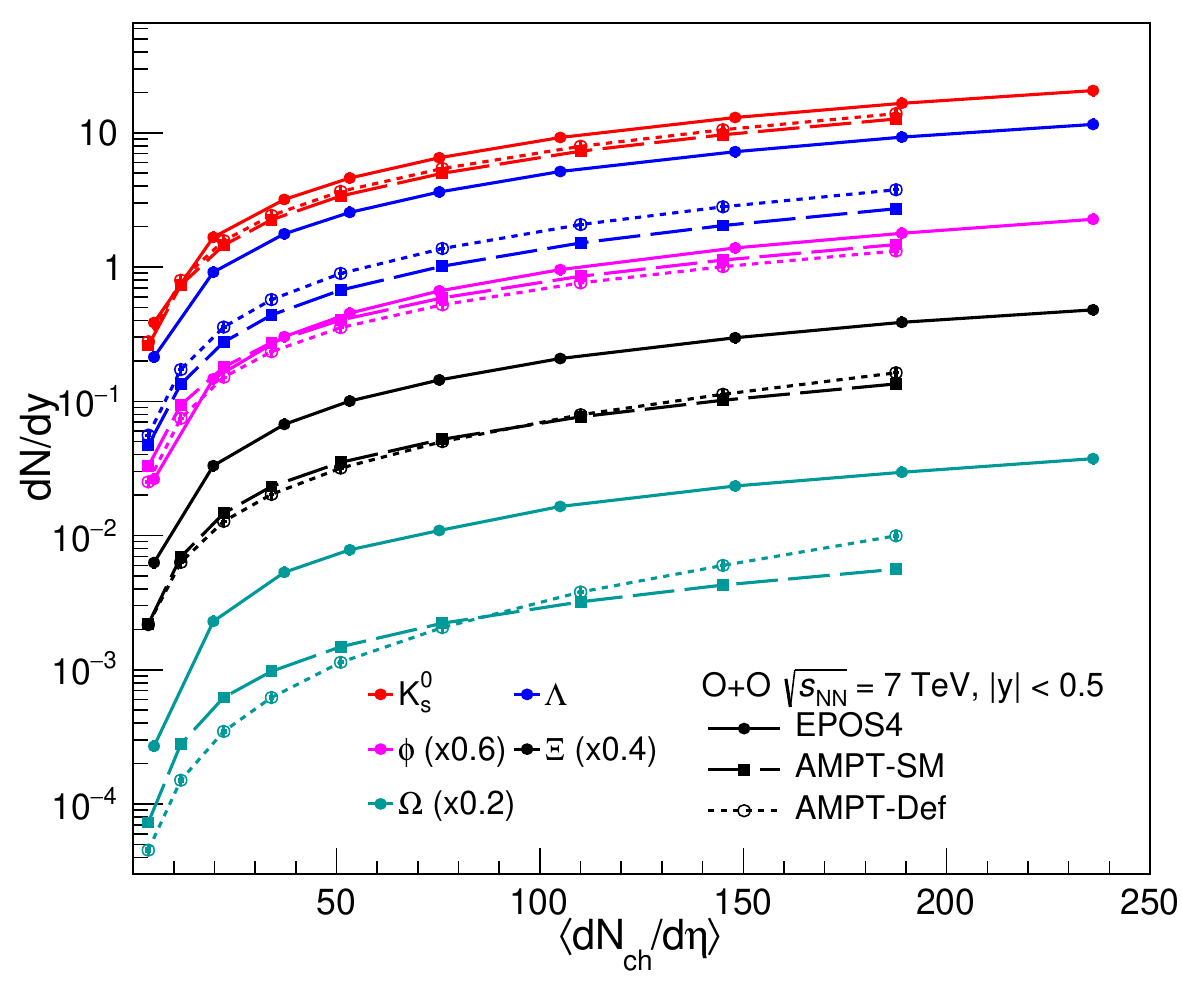}
\caption{(Color online) Multiplicity dependence of the particle yield $\mathrm{d}N/\mathrm{d}y$ of strange hadrons ({\ks}, $\Lambda$, $\Xi$, $\phi$, and $\Omega$) in \oo collisions at \sevenn using AMPT-Def, AMPT-SM and EPOS4. Solid lines are used for EPOS4, whereas the dotted and dashed lines represent AMPT-Def and AMPT-SM, respectively.}
\label{fig2}
\end{figure}

Figure~\ref{fig2} shows the predictions from the same models for the yields ($\mathrm{d}N/\mathrm{d}y$) of various strange hadrons as a function of event charged-particle multiplicity density \avdndeta. The predictions from both models show an increasing trend in the yield of strange hadrons as we move from peripheral to central collisions. It is also observed that the average yields of multi-strange hyperons decrease systematically with an increasing number of strange quarks. In EPOS4, the central collisions are always core-dominated, and the contribution of the core becomes less with decreasing centrality~\cite{Werner:2007bf}. It is observed that heavier particles show strong centrality dependence as compared to lighter particles only in peripheral collisions. In EPOS4, the strength of centrality dependence varies based on the hadron type; for instance, $\Xi$ and $\Omega$ exhibit a stronger centrality dependence compared to \ks.
The centrality dependence appears to be primarily determined by the relative significance of the corona contribution. As the contribution of corona diminishes, the yield variation with centrality becomes more pronounced~\cite{Werner:2007bf}. All models consistently predict an increase in $\mathrm{d}N/\mathrm{d}y$ with increasing \avdndeta. The predictions for $\Lambda$, $\Xi$, and $\Omega$ by EPOS4 are always larger than those from AMPT-Def and AMPT-SM. The low production of multi-strange hadrons in AMPT-SM is due to not optimized coalescence parameters in the currently used code version~\cite{He:2017tla}. 

\begin{figure}[!ht]
\includegraphics[width=0.45\textwidth]{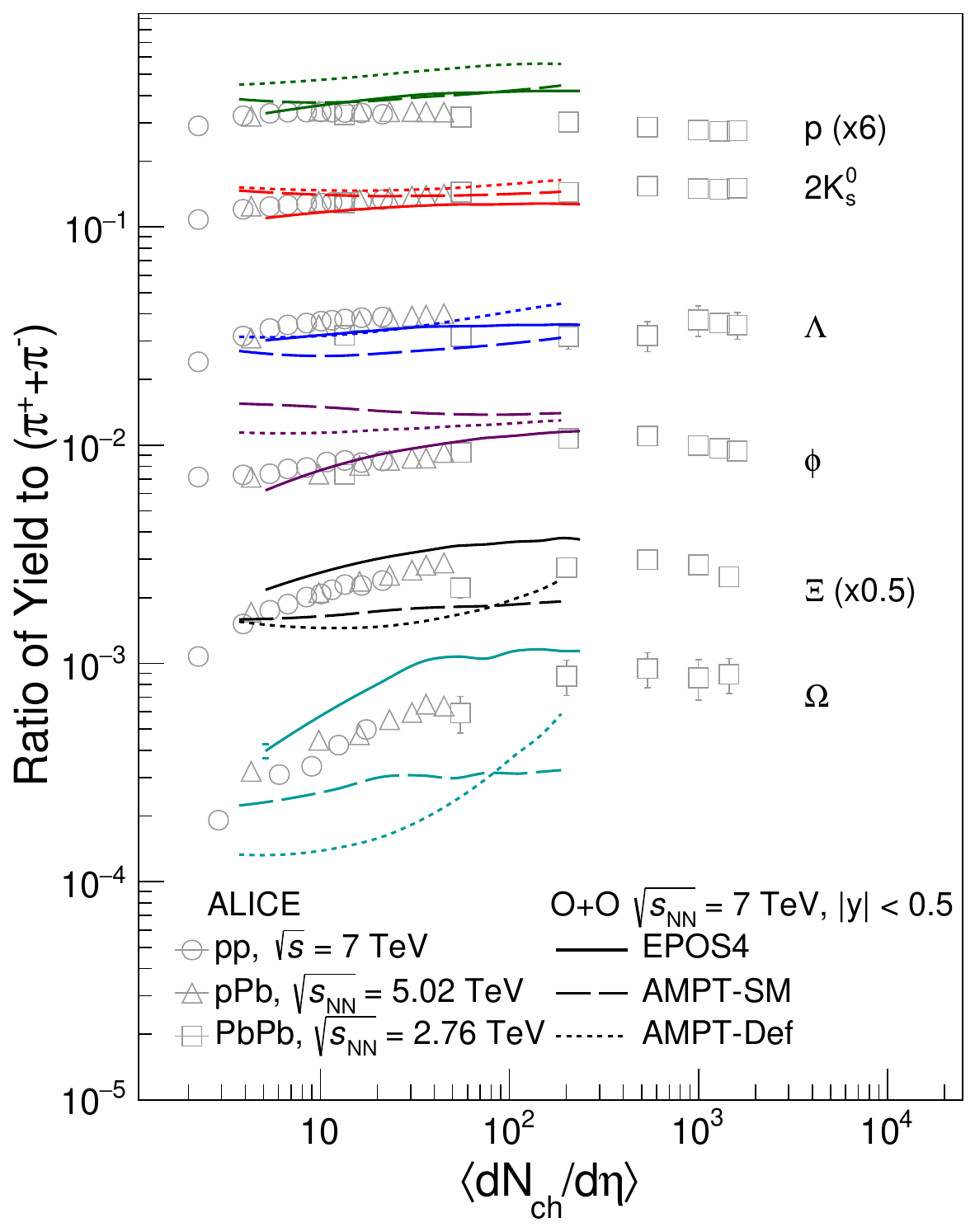}
\caption{(Color online) \ppt-integrated yield ratios of {\ks}, $\Lambda$, $\Xi$, $\phi$, and $\Omega$ to pions ($\pi^+ + \pi^-$) as a function of \avdndeta in \oo collisions at \sevenn using AMPT-Def, AMPT-SM and EPOS4. Solid lines are used for EPOS4, whereas the dotted and dashed lines represent the AMPT-Def and AMPT-SM models, respectively. The values are compared to the published results from \pp, \ppb, and \pbpb collisions~\cite{5, 45, 49, 50, 51, 52}.}
\label{fig3}
\end{figure}

To investigate the relative production of strange particles with respect to non-strange particles, the yield ratios of strange particles to pions were calculated as a function of charged particle multiplicity, and the predictions from all the models are shown in Fig.~\ref{fig3}.   
In AMPT-Def, the mechanism of string fragmentation is responsible for hadronization. This process involves the fragmentation of color strings connecting quarks and antiquarks, resulting in the production of strange hadrons. As string fragmentation does not fully account for the collective flow, it underestimates strangeness production as compared to models that include quark coalescence, like AMPT-SM~\cite{He:2017tla}. The \ppt-integrated ratios of strange hadrons over pions in the AMPT-Def are less sensitive to the multiplicity because the production of strange quarks is sensitive to fragmentation rather than collective effects. In the AMPT-SM model, the initial strange quarks originate from the melting of HIJING generated strings. The subsequent partonic phase consists solely of elastic parton-parton scatterings, which thermalize the system without producing additional strangeness. Unlike the default AMPT, hadronization occurs via quark coalescence, a mechanism that efficiently combines strange quarks into multi-strange hadrons ($\Xi$, $\Omega$) from the thermalized medium, potentially enhancing their production relative to pions, which is a signature consistent with observed QGP behavior. This enhancement exhibits a hierarchy according to strangeness content. As illustrated in Fig.~\ref{fig3}, the yield ratios of singly-strange particles ($K^0_S$, $\Lambda$) is lower in AMPT-SM than in the default model, which employs string fragmentation. These observations demonstrate the model-dependent interplay between fragmentation and coalescence mechanisms across different hadron species. AMPT simulations suggest that 0--5\% centrality in \oo collisions corresponds to a charged-particle multiplicity similar to that observed in 50--60\% centrality in \pbpb collisions.

EPOS4 includes a detailed description of the collision dynamics through viscous hydrodynamics, which is crucial for modeling the medium's collective behavior and its effect on particle production. In EPOS4, hadronization of the hydrodynamic core is described using the Cooper–Frye prescription at the freeze-out hypersurface, after which particles are mainly produced through string fragmentation. As charged-particle multiplicity increases, the ``core'' region dominates the collision. In this region, strange quarks are more likely to be produced due to the high energy density and temperature, leading to a clear strangeness enhancement, especially for multistrange hadrons ($\Xi$ and $\Omega$) in events with large charged-particle multiplicity.
In low-multiplicity events, the corona region of EPOS4, where hard processes such as jet fragmentation dominate, generates a smaller number of strange quarks. Consequently, the ratios of strange hadrons relative to pions are lower than in high-multiplicity events, where the core region dominates and strangeness is enhanced. The predictions of the yield ratios relative to pions in \oo collisions from all models are compared with those measured in \pp, \ppb, and \pbpb collisions at available LHC energies~\cite{5, 45, 49, 50, 51, 52}. Interestingly, these predictions from all models show a clear final-state multiplicity overlap with \pp, \ppb, and \pbpb collisions. Extensive studies of strangeness production with EPOS4 and AMPT across different collision systems have been reported in Refs.~\cite{40, Werner:2023mod, 42, He:2017tla, Shao:2020sqr, Sahoo:2020jsa}.

\begin{figure}[t!]
    \centering
    \begin{subfigure}
        \centering
        \includegraphics[width=0.49\linewidth]{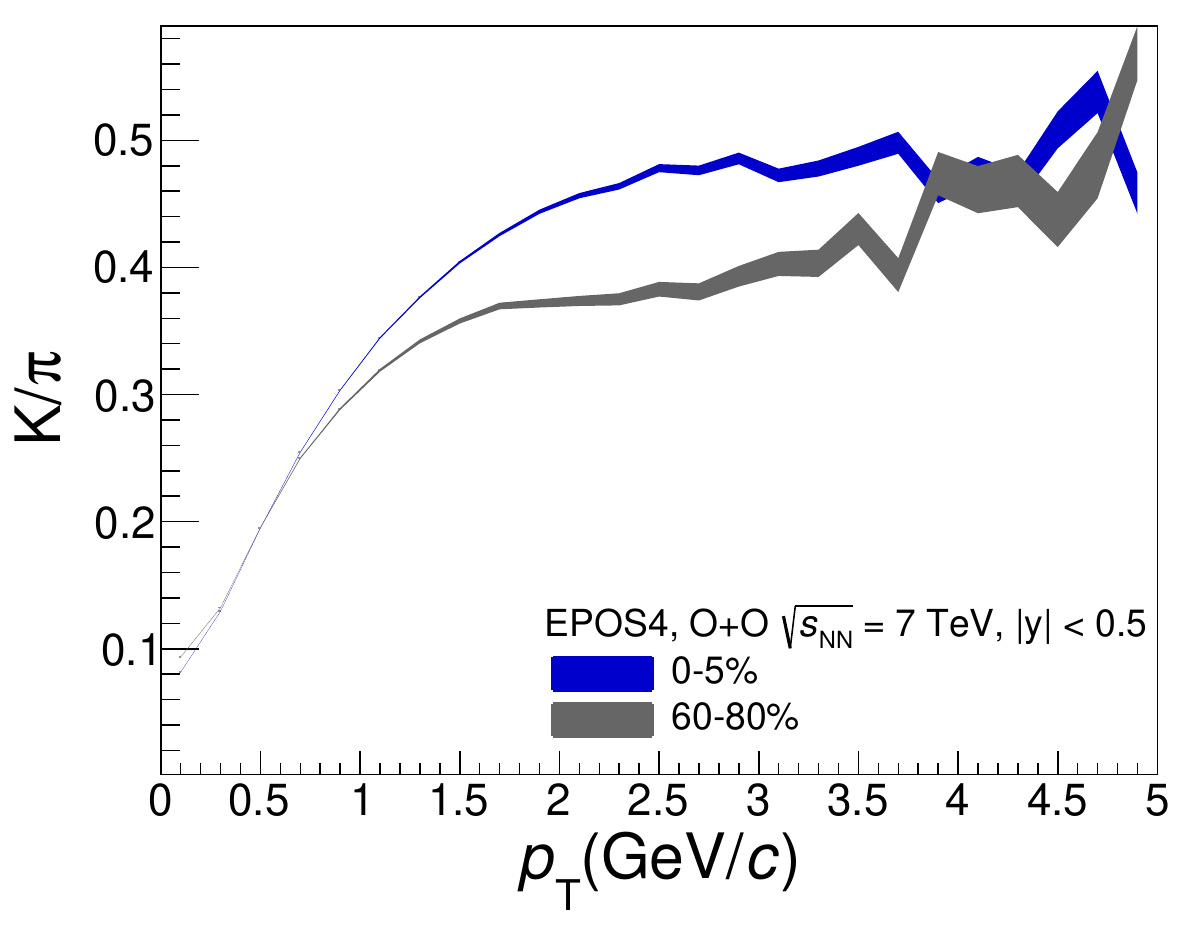} 
        \includegraphics[width=0.49\linewidth]{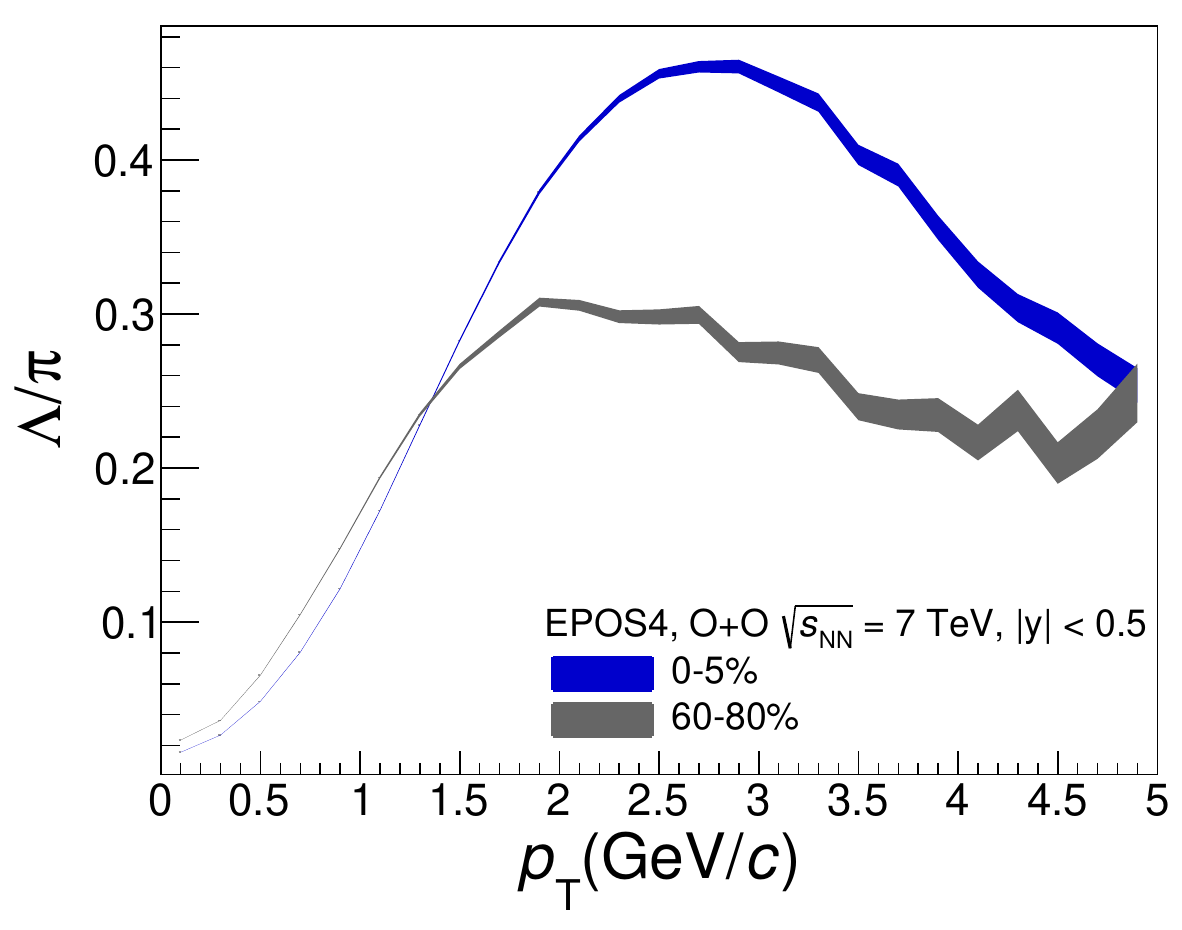}
        \includegraphics[width=0.49\linewidth]{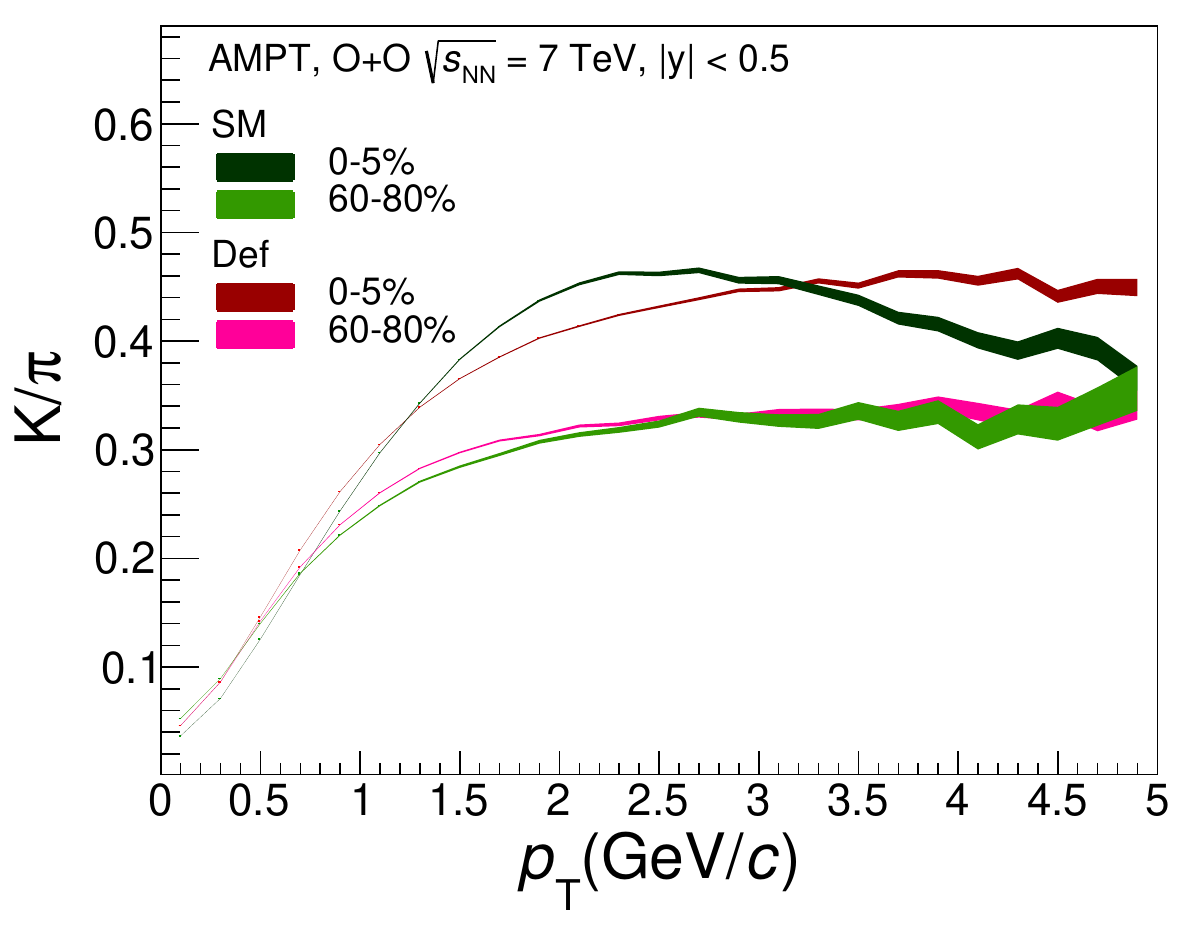}
        \includegraphics[width=0.49\linewidth]{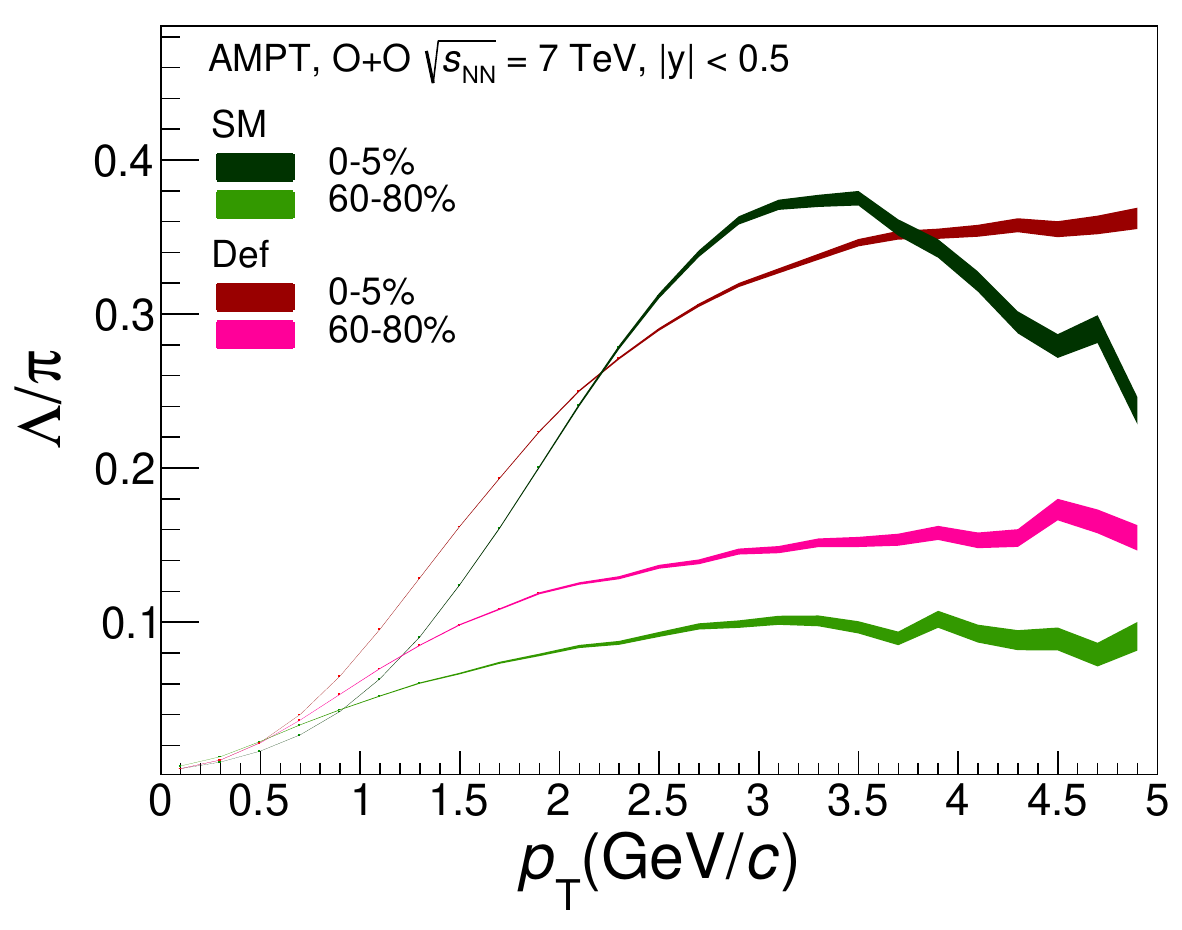}
    \end{subfigure}%
    \caption{(Color online) \ppt-differential $K/\pi$ and $\Lambda/\pi$ ratios in most central (0--5\%) and the most peripheral (60--80\%) \oo collisions at \sevenn using EPOS4 (top) and AMPT (bottom). Different colors are used to represent different models across various centrality classes. 
    }\label{fig4}
\end{figure}

Figure~\ref{fig4} shows the \ppt-differential particle ratios ($K/\pi$ and $\Lambda/\pi$) in the most central (0--5\%) and the most peripheral (60--80\%) \oo collisions at \sevenn using all models. Particle ratios involving different hadron species serve as a direct probe of the relative abundances and interaction dynamics of the underlying quark constituents within the hot and dense medium. Both ratios serve as a measure of the strangeness enhancement and are observed for both EPOS4 and AMPT models at intermediate \ppt. This enhancement shows weak centrality dependence at low \ppt while exhibiting strong centrality dependence in the intermediate-\ppt region. The production of strangeness is maximum for the central collisions at intermediate-\ppt, while it decreases with centrality. The $\Lambda/\pi$ ratio exhibits similar behavior due to similar strangeness content in $K$ and $\Lambda$. 
Additionally, the $\Lambda/\pi$ ratios from EPOS4 and AMPT-SM show a clear peak around 2--3 GeV/$c$, which is more pronounced in central collisions. In AMPT-SM, this peak arises from baryon recombination via quark coalescence, resulting in enhanced baryon production at intermediate \ppt~\cite{Fries:2008hs}. AMPT-Def is based on string fragmentation without a partonic phase, lacks a pronounced peak and serves as a baseline for vacuum-like hadronization. In EPOS4, this peak structure most likely arises from the complex interplay between the hydrodynamically expanding core and the corona. The strangeness enhancement in the $K/\pi$ ratio at intermediate \ppt in EPOS4 shows qualitative agreement with AMPT-SM.

\begin{figure}[h]
    \centering
    \begin{subfigure}
        \centering
        \includegraphics[width=0.49\linewidth]{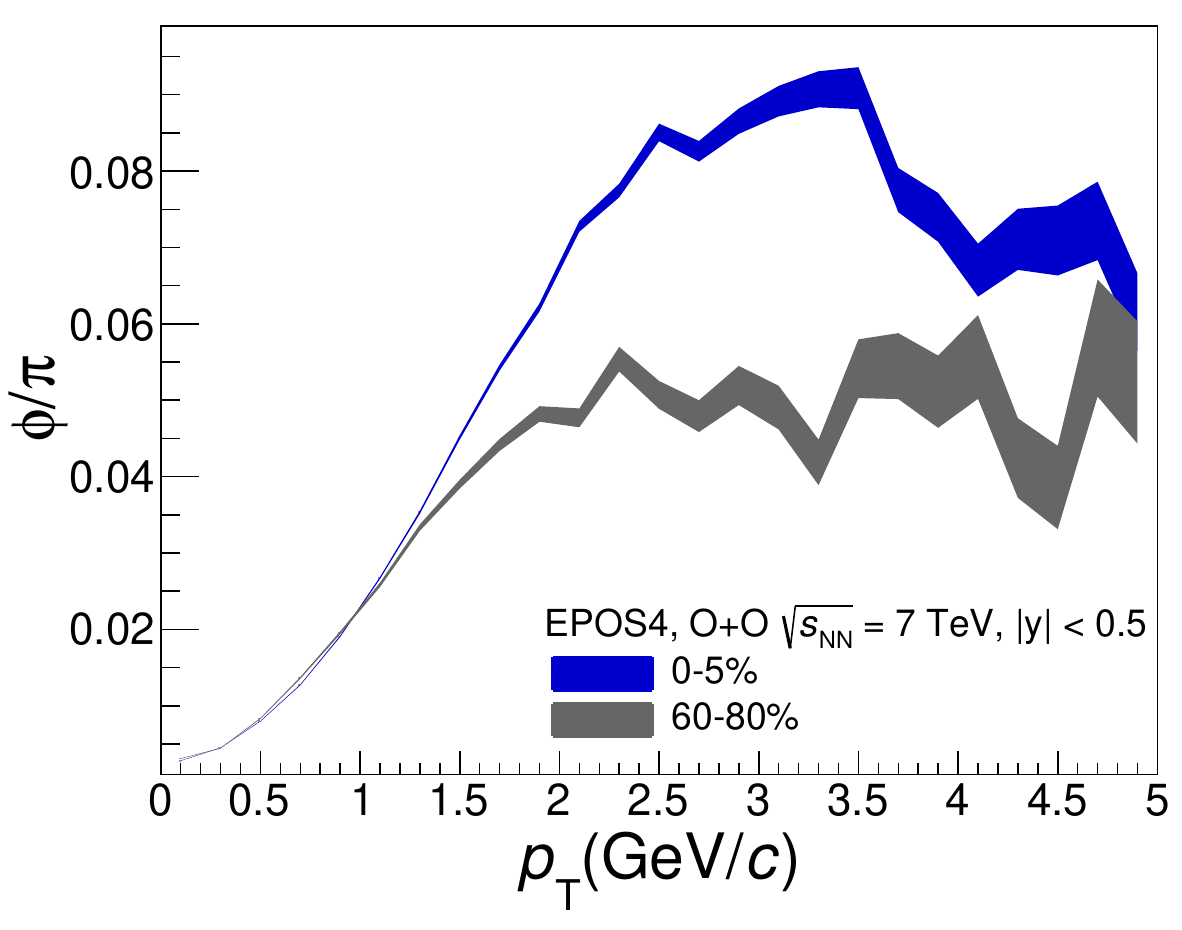}
        \includegraphics[width=0.49\linewidth]{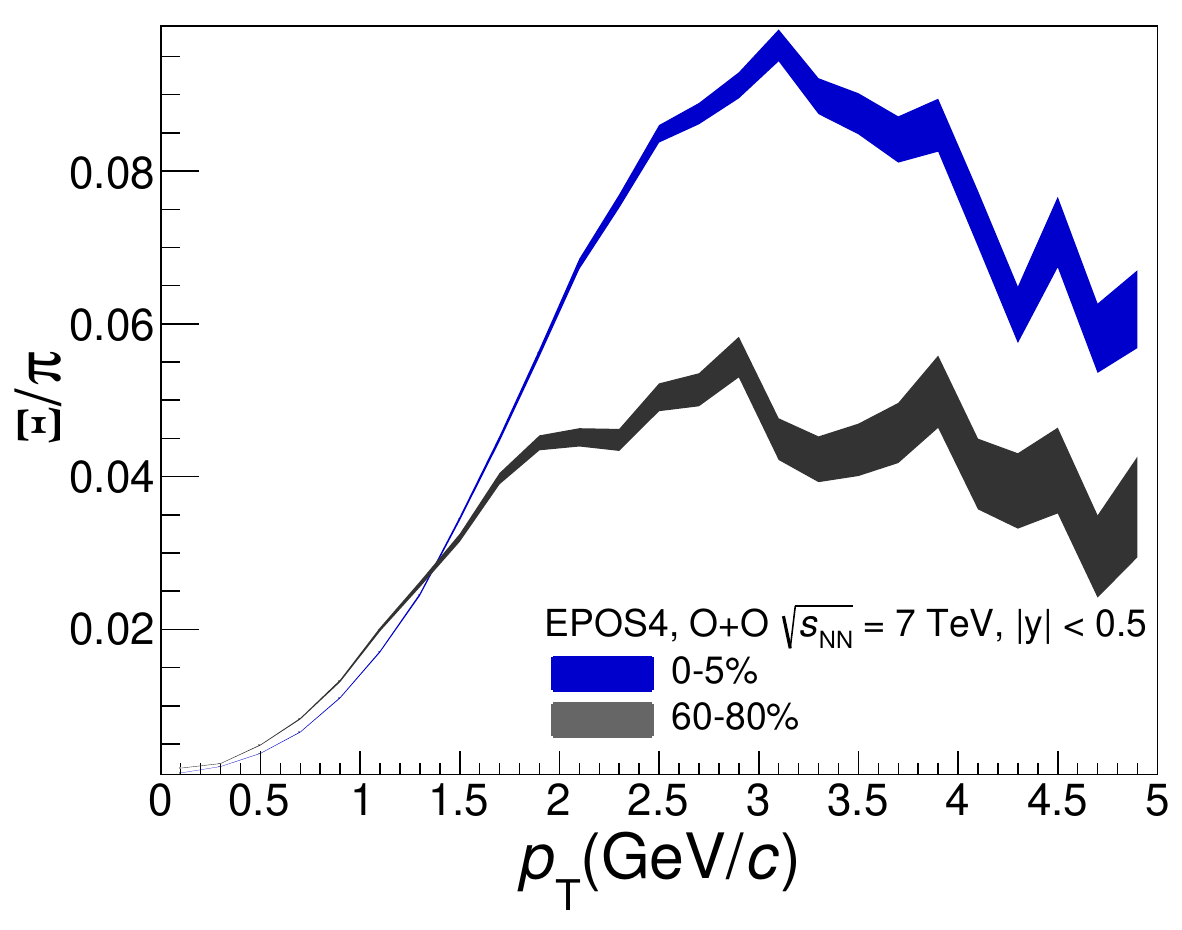}
        \includegraphics[width=0.49\linewidth]{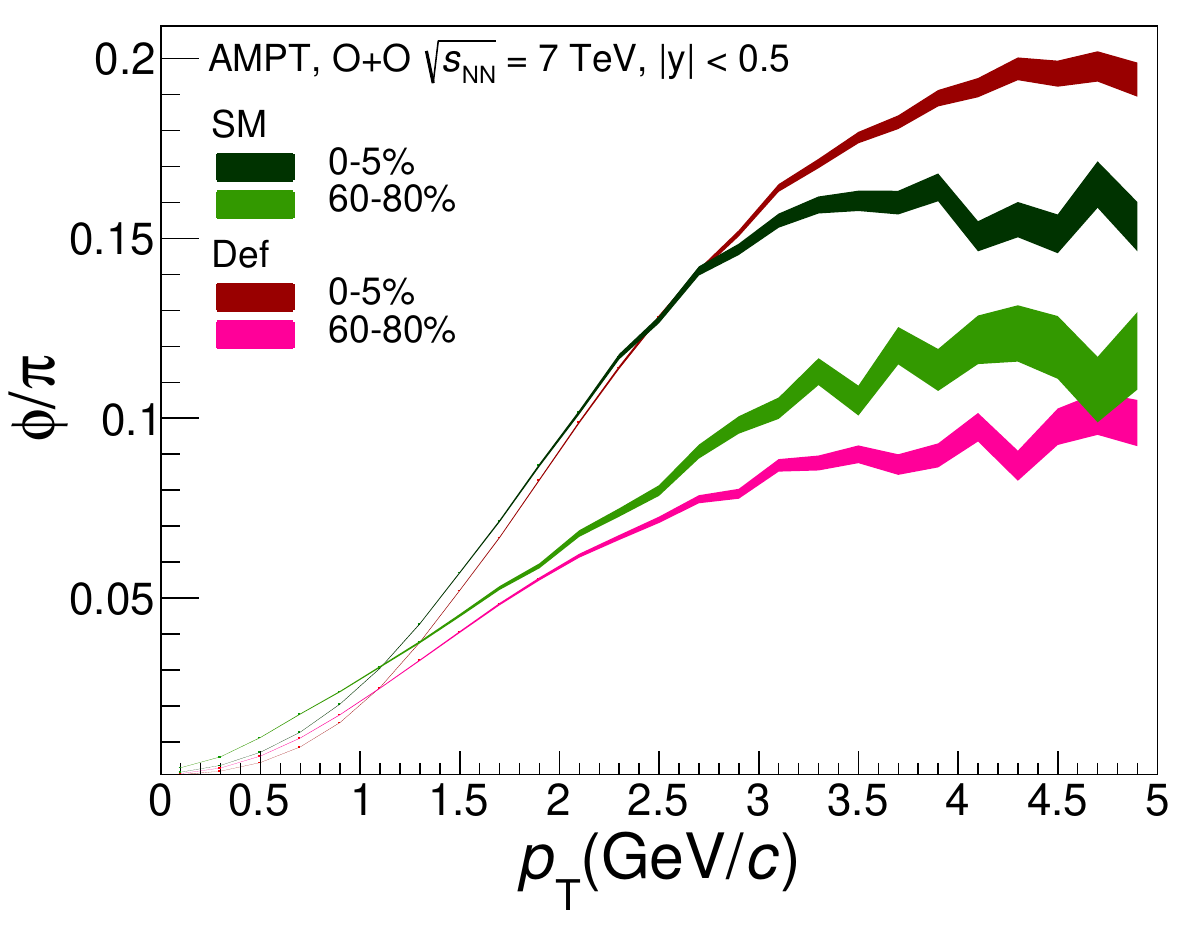}      
        \includegraphics[width=0.49\linewidth]{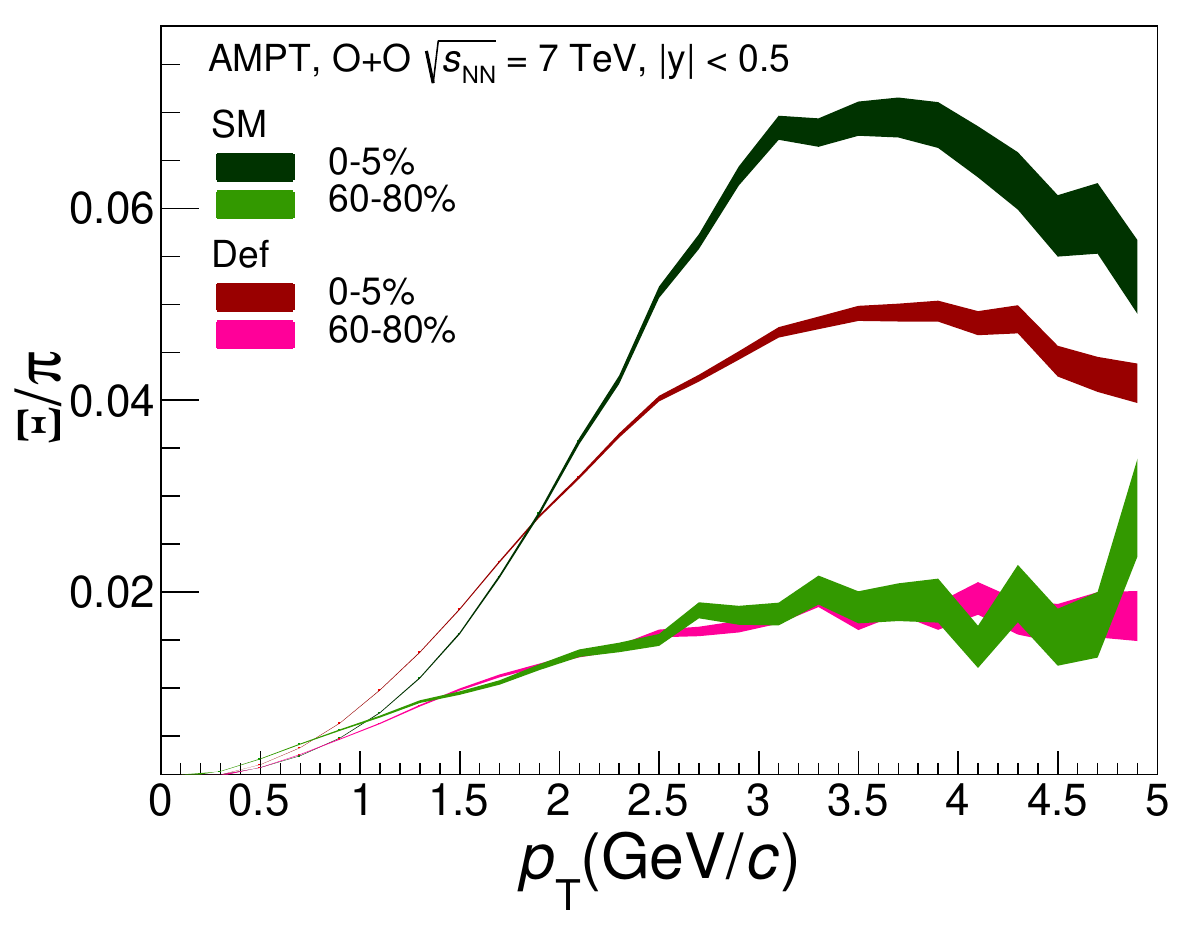}
    \end{subfigure}%
    \caption{(Color online) \ppt-differential $\phi/\pi$ and $\Xi/\pi$ ratios in most central (0--5\%) and the most peripheral (60--80\%) \oo collisions at \sevenn using EPOS4 (top) and AMPT (bottom). Different colors are used to represent different models across various centrality classes. 
    }\label{fig5}
\end{figure}

Figure~\ref{fig5} shows the $\phi/\pi$ and $\Xi/\pi$ ratios in the most central (0--5\%) and peripheral (60--80\%) \oo collisions at \sevenn using EPOS4, AMPT-Def, and AMPT-SM models. The rise in $\phi/\pi$ and $\Xi/\pi$ ratios is more gradual, exhibiting a rapid increase at low \ppt as compared to $K/\pi$ and $\Lambda/\pi$ ratios. Both ratios show weak centrality dependence at low \ppt but strong centrality dependence at intermediate-\ppt from all models. It is observed that at intermediate \ppt, the $\phi/\pi$ ratio is twice as large for AMPT-SM as compared to EPOS4 in the most central collisions. This may be because, at the intermediate-\ppt region where the momentum of a particle becomes comparable or more than the mass of the strange quark, the production probability of $s\bar s$ pairs increases significantly. Additionally, $\phi$ mesons are not only produced from the fragmentation of excited strings in the initial collisions but also from the hadronic matter through various baryon-baryon, meson-baryon, and meson-meson scatterings. This leads to a corresponding rise in the observed $\phi/\pi$ ratio at intermediate \ppt. The difference in $\phi/\pi$ ratios in AMPT-SM and EPOS4 might also be attributed to the influence of the Lund string fragmentation parameters on the yields of $\phi$ mesons. AMPT-SM may have implemented these parameters more effectively, leading to a stronger preference for strange quark hadronization than in EPOS4~\cite{Tiwari:2020ult}. EPOS4 does not include a coalescence mechanism, and the $\phi/\pi$ ratio reflects the complex interplay between core-corona separation, hydrodynamics, collective flow, and particle production mechanisms across different \ppt ranges. $\Xi/\pi$ ratios also show strong centrality dependence at the intermediate \ppt in all models. Similar to the $\Lambda/\pi$ ratios, the $\Xi/\pi$ ratios exhibit a pronounced peak with a maximum of around 2--3 GeV/$c$ in most central collisions.

\begin{figure}[h]
    \centering
    \begin{subfigure}
        \centering
         \includegraphics[width=0.49\linewidth]{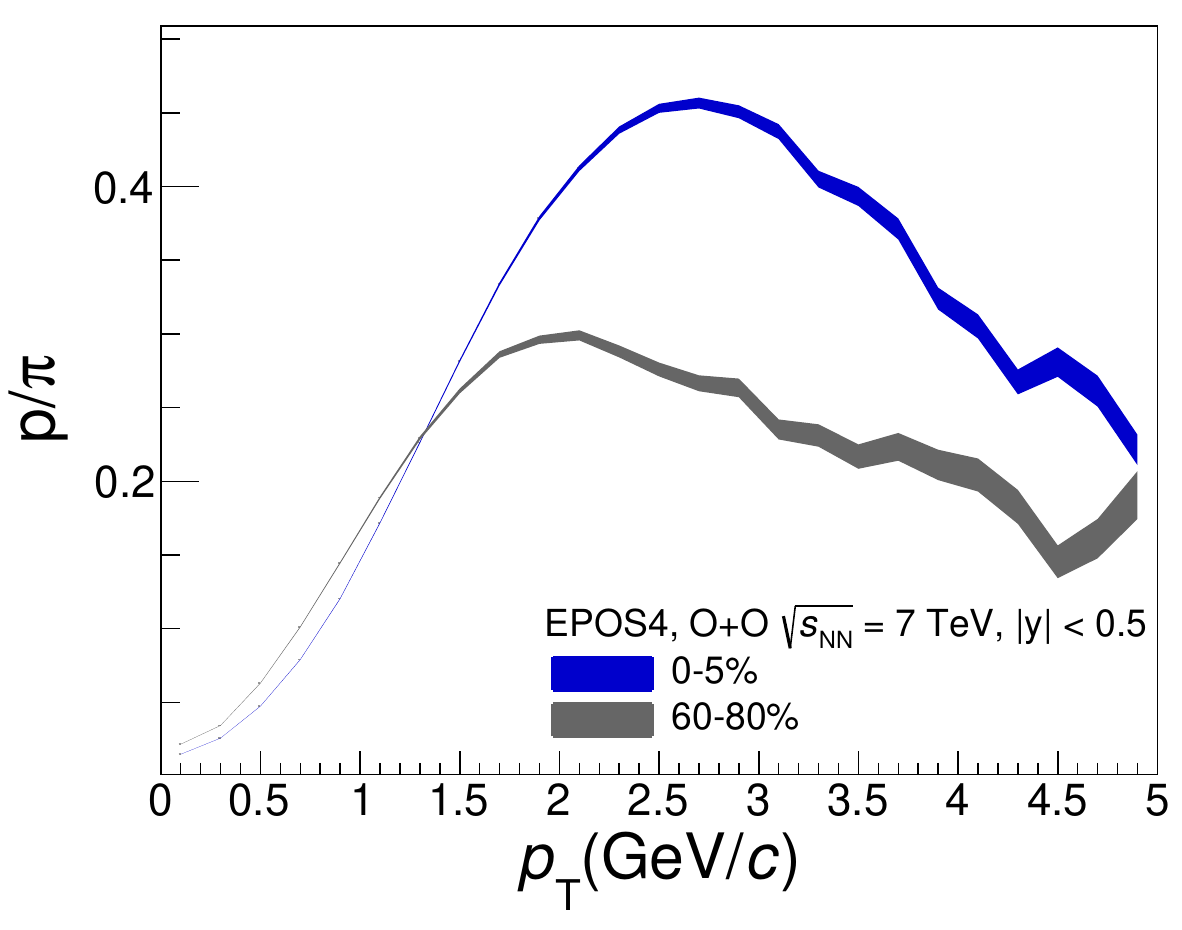}
         \includegraphics[width=0.49\linewidth]{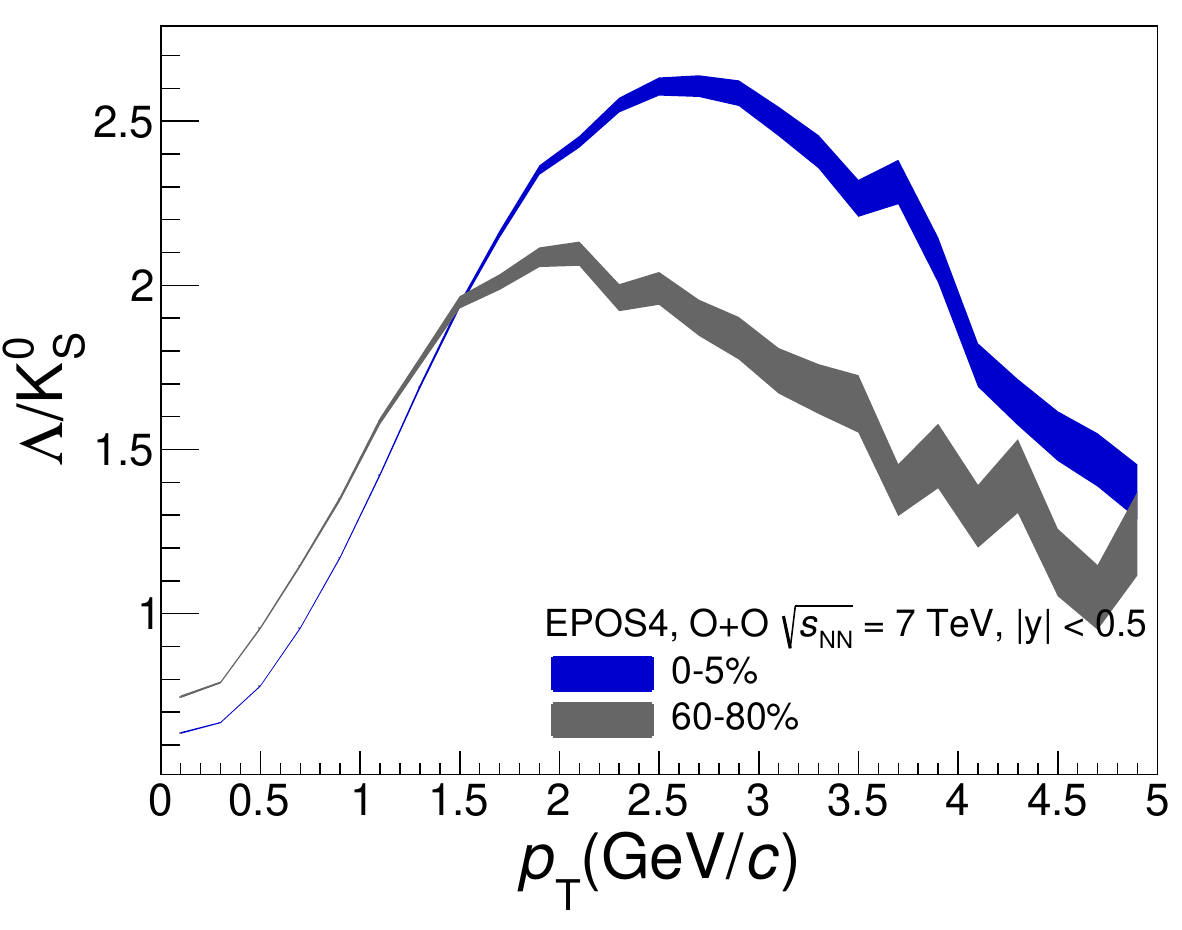}   
         \includegraphics[width=0.49\linewidth]{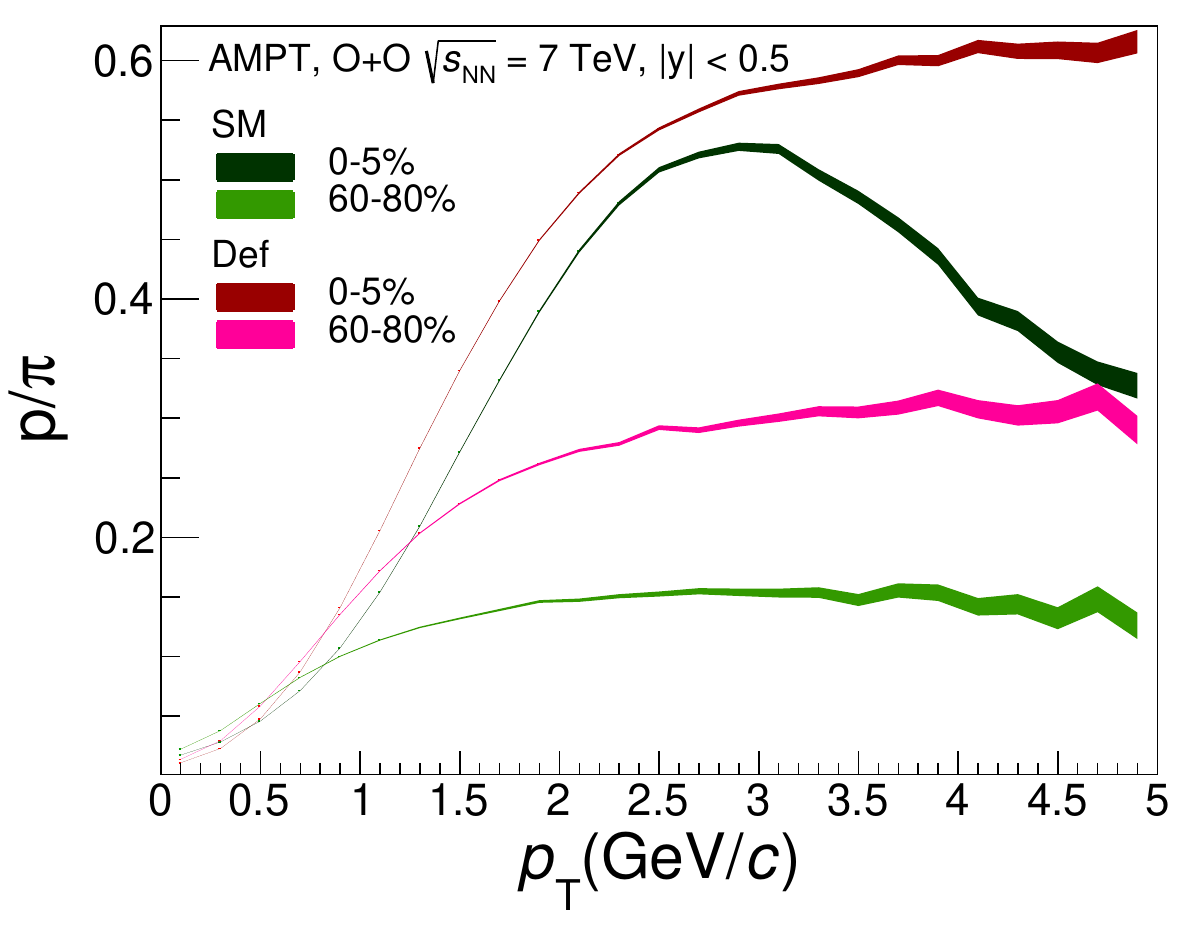}
         \includegraphics[width=0.49\linewidth]{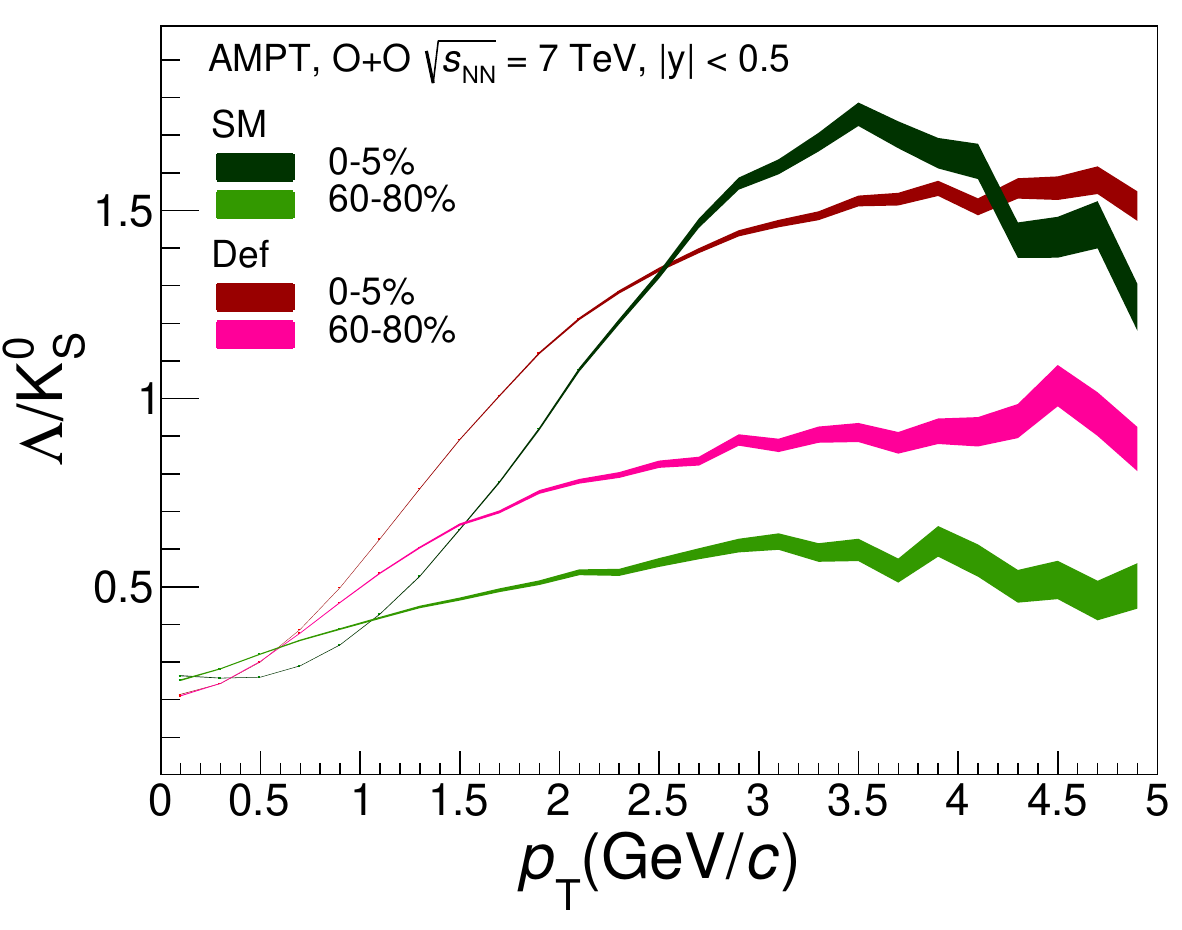}
    \end{subfigure}%
    \caption{(Color online) \ppt-differential baryon-to-meson ($p/\pi$ and $\Lambda$/{\ks}) ratios in most central (0--5\%) and the most peripheral (60--80\%) \oo collisions using EPOS4 (top) and AMPT (bottom). Different colors are used to represent different models across various centrality classes. 
    }
    \label{fig6}
\end{figure}

Figure~\ref{fig6} shows \ppt-differential $p/\pi$ and $\Lambda$/{\ks} in the most central (0--5\%) and peripheral (60--80\%) \oo collisions at \sevenn using all models. The \ppt-differential $p/\pi$ (the lightest baryon to the lightest meson) ratio serves as a proxy of the relative production of baryons compared to mesons. Both ratios show a strong centrality dependence at intermediate \ppt. A similar peak around 2--3 GeV/$c$ in most central collisions is observed in $p/\pi$ ratios from EPOS4 and AMPT-SM at intermediate \ppt. 

In AMPT, we observe a peak at intermediate \ppt in central collisions, while a plateau appears at intermediate \ppt in peripheral collisions similar to that observed in \pp, \ppb, and \pbpb collisions at the LHC~\cite{5, 45, 49, 50, 51, 52}. The $\Lambda$/{\ks} ratios also exhibit strong centrality dependence at intermediate \ppt.

The enhancement in $\Lambda$/{\ks} in 0--5\% central collisions from EPOS4 is significantly larger than that predicted by the AMPT-SM model despite having a similar distribution. This discrepancy may be attributed to the strong radial flow implemented in EPOS4, which affects baryons like $\Lambda$ at higher multiplicities. In contrast, AMPT incorporates a certain degree of flow effects. This result aligns closely with the observed collision dynamics at this center-of-mass energy, indicating that the EPOS4 treatment of strangeness production mechanisms, especially for these particles, shows better consistency with the existing experimental data~\cite{42, Werner:2013tya}. 

\section{Conclusions}
\label{sec4}

We present predictions of various observables for (multi)strange hadrons ({\ks}, {\lam}, {\xii}, $\phi$, and {\om}) in \oo collisions at \sevenn using the recently updated hydrodynamic-based EPOS4 and two different versions of the AMPT model. These models have been utilized to investigate the transverse momentum (\ppt) spectra, multiplicity dependence of the particle yield $\mathrm{d}N/\mathrm{d}y$, yield ratio of various strange hadrons, and \ppt-differential ratios of different strange meson and baryon relative to pions.
EPOS4 performs well in predicting the enhancement of strangeness, while the current version of AMPT does not. None of the models can fully describe trends of strangeness enhancement observed in existing experimental data from other collision systems. A clear final-state multiplicity overlap is observed with \pp, \ppb, and \pbpb collisions from all the models. The \ppt-differential strange hadron to pion ratios show a clear centrality dependence. AMPT predicts a stronger enhancement in the $\phi/\pi$ ratio compared to EPOS4 due to the effective implementation of the Lund string fragmentation parameters. The stronger enhancement of the $\Lambda$/{\ks} ratio in the intermediate-\ppt region in EPOS4 compared to AMPT can be attributed to the inclusion of a full hydrodynamic evolution in EPOS4, combined with its hadronization treatment (string fragmentation). Both factors enhance baryon production and lead to better agreement with experimental data from other collision systems. The upcoming data on \oo collisions at the LHC will help constrain and refine the model parameters.

\section{Acknowledgements}
The authors would like to thank W.~J.~Llope and S.K. Grossberndt for careful reading and discussions. H.~R. is supported in part by the National Science Foundation (NSF) within the framework of the JETSCAPE collaboration, under grant numbers ACI-1550300 OAC-2004571 (CSSI:X-SCAPE) and in part by the U.S. Department of Energy (DOE) under grant number DE-SC0021969.

\bibliographystyle{utphys}
\bibliography{bib}

\end{document}